\documentclass[aps,prx,twocolumn,superscriptaddress,nofootinbib,notitlepage,longbibliography]{revtex4-1}
\usepackage{bm}
\usepackage{hhline}
\usepackage{amsmath}
\usepackage{amssymb}
\usepackage{mathdots}
\usepackage{tabularx}
\usepackage{graphicx}
\usepackage{hyperref}
\usepackage{bbding}
\usepackage{tikz}
\usepackage{bbding}
\usepackage{braket}
\usepackage{bm}
\usepackage{multirow}

\begin{document}
	\title{Dynamical Generation of Rectified Electric Current}
\author{Jin-Xin Hu}
\email{jhuphy@ust.hk}
\affiliation{Department of Physics, Hong Kong University of Science and Technology, Clear Water Bay, Hong Kong, China}
\author{Congjun Wu}
\email{wucongjun@westlake.edu.cn}
\affiliation{New Cornerstone Science Laboratory, Department of Physics, School of Science, Westlake University, Hangzhou 310024, Zhejiang, China}
\affiliation{Institute of Natural Sciences, Westlake Institute for Advanced Study, Hangzhou 310024, Zhejiang, China}
\affiliation{Institute for Theoretical Sciences, Westlake University, Hangzhou 310024, Zhejiang, China}
\affiliation{Key Laboratory for Quantum Materials of Zhejiang Province, School of Science, Westlake University, Hangzhou 310024, Zhejiang, China}

\date{\today}
\begin{abstract}
Rectification is a fundamental nonlinear transport process that converts an alternating drive into a direct current. In this work, we propose a general theoretical framework for electric current rectification triggered by a dynamical external drive that couples to an arbitrary well-defined operator of a periodic system, and which in the static limit forbids any steady current. In the dynamical regime, the finite frequency $\Omega$ of the time-varying drive breaks time-translation invariance and injects energy into the system, enabling a second-order {\it nonlinear rectified} current that has no static counterpart. This rectification process has two distinct origins: (i) an impurity-scattering-modified distribution function at finite frequency, and (ii) a time‑domain anomalous velocity stemming from a dynamical mixed Berry curvature. Both contributions persist when the driving frequency lies well below the optical transition gap. Applying our general theory to a buckled magnetic system subject to an out‑of‑plane oscillating electric field, we characterize the generated current as a {\it nonlinear magnetoelectric gyrotropic effect} and predict that the induced rectified current is sensitive to the magnetic order, thereby offering a feasible electrical probe of N\'{e}el order in non-coplanar antiferromagnets.
\end{abstract}
\maketitle

\emph{Introduction.}---The conversion of alternating electromagnetic fields into direct electric currents, known as rectification, is a phenomenon of fundamental importance in both condensed matter physics and modern electronics~\cite{schaefer1965rectifier,buttiker1993capacitance,wei1997current,siwy2006ion,ono2002current,isobe2020high,ideue2017bulk}. Rectification can be viewed fundamentally as a nonreciprocal transport response, where the current–voltage relation lacks symmetry under reversal of the applied bias. Traditionally, rectifiers are implemented with semiconductor devices, such as diodes, which exploit the asymmetric conduction characteristics of a $p$-$n$ junction~\cite{shockley1949theory,streetman2000solid}. While these conventional rectifiers rely on macroscopic circuit elements and junctions, in quantum materials rectification can arise from purely microscopic mechanisms driven by the oscillating optical field of light or low-frequency AC voltages. At optical frequencies, the bulk photogalvanic effect generates a DC photocurrent in noncentrosymmetric crystals under uniform irradiation~\cite{dai2023recent,ma2023photocurrent,von1981theory,aversa1995nonlinear,sipe2000second,morimoto2016topological,ahn2020low,kaplan2020nonvanishing,holder2020consequences,ma2021topology,watanabe2021chiral,ahn2022riemannian,shi2021geometric,matsyshyn2021rabi,nagaosa2017concept,shi2021geometric}. At low frequencies, rectification in quantum materials can originate from quantum geometry, including Berry curvature~\cite{sodemann2015quantum,matsyshyn2019nonlinear,kumar2021room,du2021nonlinear,ma2019observation,du2021quantum,zhang2021terahertz,onishi2024high,huang2023giant,hu2022nonlinear} and quantum metric~\cite{wang2021intrinsic,liu2021intrinsic,gao2023quantum,wang2023quantum,das2023intrinsic,kaplan2024unification,qiang2026clarification}.

\begin{figure}
\includegraphics[width=1.0\linewidth]{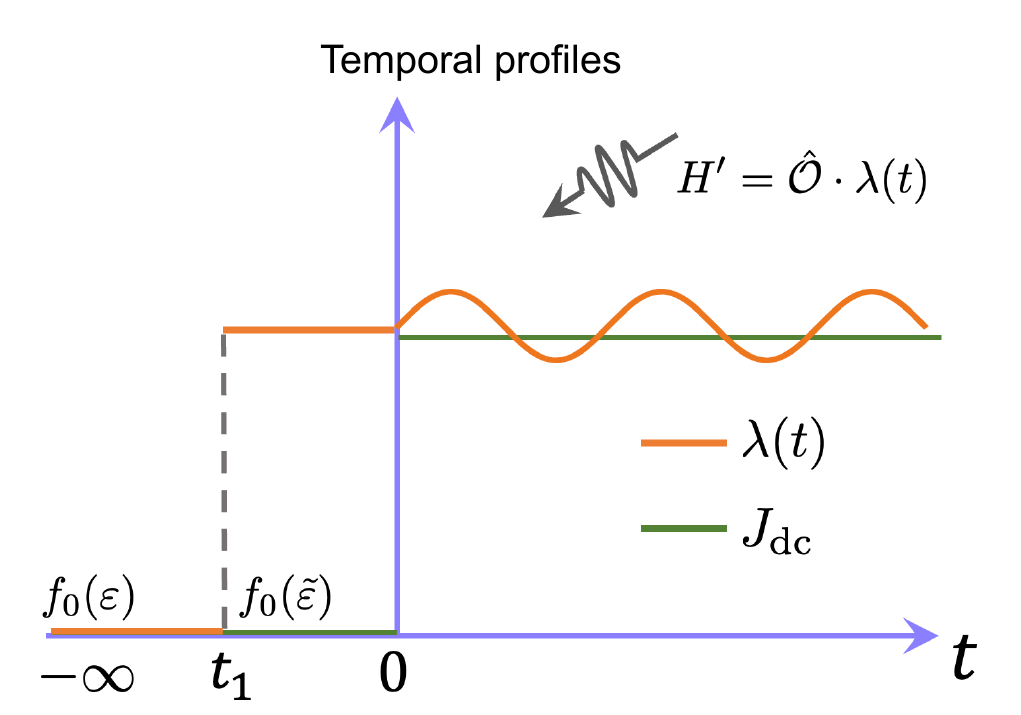}
\caption{Schematic illustration of rectified current generation by a dynamical drive $\lambda(t)$ coupled to a local operator $\hat{\mathcal{O}}$. At times $t < 0$, the system is in equilibrium. The drive is adiabatically turned on during the interval $-\infty < t < t_1$, reaching a finite amplitude with zero frequency for $t_1<t<0$. While a static component of the drive would merely redistribute carriers, leading to a modified steady-state distribution, the finite frequency $\Omega$ at $t > 0$ enables the generation of a rectified DC current $J_{\mathrm{dc}}$.}
\label{fig:fig1}
\end{figure}

Transport phenomena induced by dynamical external drives that couple to the internal degrees of freedom of Bloch electrons have attracted intense recent interest~\cite{zhai2025twistronics,jia2026nonlinear,hu2025nonlinear,hu2026orbital,xiang2025spin}. For comparison, a static electric field parallel to the plane of electron motion accelerates the electrons, leading to a steady current in the presence of impurity scattering~\cite{mahan2013many}. In contrast, a static magnetic field cannot generate a steady current, as it performs no work on the electrons and preserves time-translation invariance~\cite{yamamoto2015generalized}. In chiral crystals, for example, the gyrotropic magnetic effect can generate an oscillating charge current in response to a oscillating magnetic field~\cite{zhong2016gyrotropic,wang2020optical,paul2025gyrotropic,xiang2025intrinsic}. A crucial feature of this effect is the coupling of the magnetic field to the orbital or spin magnetic moments of electrons. As the frequency of the drive is reduced to zero, the system adiabatically follows the instantaneous Hamiltonian, returning to its static equilibrium state, thereby restoring time-translation invariance and eliminating any macroscopic current~\cite{xiao2025proper}. Generating a finite DC current in such systems thus requires a mechanism that explicitly prevents this adiabatic restoration.

Major attention and existing theories of current generation driven by such dynamical external fields have mainly focused on first-order AC responses. In this work, we demonstrate that a dynamical external driving field, which preserves time-translation invariance in the static limit, can generate a rectified DC electric current even when the driving frequency lies well below the optical gap. Crucially, this current arises as a second-order response to the perturbation and exists only at finite frequencies, vanishing identically in the static limit. This behavior reveals a fundamentally distinct mechanism for rectified current generation. By developing a time-dependent nonlinear transport theory, we identify both extrinsic (scattering process) and intrinsic (quantum geometric) contributions to this second-order dynamical rectification. As a concrete and physically feasible implementation, we apply our theory to a buckled magnetic system subject to an out-of-plane oscillating electric field that couples to the vertical electric dipole moment. We further predict that the resulting rectified current is sensitive to N\'{e}el ordering. This scheme also paves the way toward magnetic memory devices in antiferromagnetic spintronics.

\emph{Time-dependent nonlinear transport theory.}---We adopt the semiclassical framework to study the dynamical current generation. Without loss of generality, we consider a Bloch electron perturbed by a time-varying dynamical drive $\bm{\lambda}(t)$ (see Fig.~\ref{fig:fig1}). At $t=-\infty$, the system is in equilibrium, described by the distribution function $f_0(\varepsilon)$. Under the external perturbation, it reaches the distribution $f_0(\tilde{\varepsilon})$ at $t=t_1$. For $t>0$, we assume that the drive varies slowly enough that the system evolves adiabatically, remaining in an instantaneous eigenstate of the time-dependent Hamiltonian
\begin{equation}
H=H_0+\hat{\bm{\mathcal{O}}}\cdot\bm{\lambda}(t),
\end{equation}
where the drive can be written as $\bm{\lambda}(t)=2\bm{\lambda}\cos(\Omega t)$, with the frequency $\Omega$ lying below the interband transition threshold ($\hbar\Omega \ll \Delta_{\mathrm{gap}}$). In this regime, the electric current density can be expressed in terms of the time-varying electron velocity $\bm{v}(\bm{k},t)$ and distribution function  $f(\bm{k},t)$ as 
\begin{equation}
\bm{J}(t)=-e\sum_n\int_{\bm{k}}\bm{v}_n(\bm{k},t)f_n(\bm{k},t).
\end{equation}
Here $n$ is the band index in a crystal and $\int_{\bm{k}}=\int d\bm{k}/(2\pi)^d$ with $d$ the dimension. Under the oscillating field with finite $\Omega$, the electrons are in the excited state with the new distribution function $\tilde{f}_n(\bm{k},t)$. Within the relaxation-time approximation, the semiclassical Boltzmann equation reads
\begin{equation}
\partial_t \tilde{f}_{n}(\bm{k},t)=-\Gamma\{\tilde{f}_{n}(\bm{k},t)-f[\tilde{\varepsilon}_n(\bm{k},t)]\}.
\end{equation}
Here we neglect the momentum and position dependent on the distribution function. $\Gamma=1/\tau$ is the scattering rate that describes the system returning to the instantaneous equilibrium state $f[\tilde{\varepsilon}_n(\bm{k},t)]$. We notice that the time-dependent band energy is expanded to $\tilde{\varepsilon}_n(\bm{k},t)=\varepsilon_n(\bm{k})+\mathcal{O}_{n}^a\lambda^a(t)$. We have $\tilde{f}_{n}(\bm{k},t)=f_{n}^N(\bm{k},t)+f[\tilde{\varepsilon}_n(\bm{k},t)]$, and to the $N$-th order expansion, we can solve $f_{n}^N(\bm{k},t)$ as
\begin{equation}
f_{n}^N(\bm{k},t)=2\mathrm{Re}[\frac{N\Omega\tau}{-N\Omega\tau+i}\frac{1}{N!}f^N_{n}(\mathcal{O}_{n}\cdot \bm{\lambda})^Ne^{iN\Omega t
}].
\end{equation}
The impurity scattering modified distribution function explicitly depends on the relaxation time $\tau$. As we will see later, this term gives the extrinsic recitfied current. Next we derive the time-varying band velocity $\bm{v}_n(\bm{k},t)$ by using the time-dependent Schrieffer-Wolff transformation~\cite{kaplan2024unification,fang2024quantum} (more details, see Appendix A). The time-varying band velocity is give by
\begin{equation}
v^b_{n}(\bm{k},t)=\partial_b\tilde{\varepsilon}_{n}(\bm{k},t) +\lambda^a [\mathcal{F}_\lambda]_{n}^{ab}(t).
\end{equation}
Obviously, $v^b_{n}(\bm{k},t)$ is dressed by the band energy correction and an anomalous velocity with the time-dependent mixed Berry curvature 
\begin{equation}
\label{eq:eq_mbc}
[\mathcal{F}_\lambda]_{n}^{ab}(t)=2\mathrm{Im}\sum_{\substack{m\neq n\\ \omega=\pm\Omega}}\mathcal{O}_{nm}^av_{mn}^b \frac{-i\omega e^{i\omega t}}{\varepsilon_{nm}(\varepsilon_{nm}+\omega)}.
\end{equation}
As seen, any dynamical drive ($\lambda^a$, along $a$ direction) generates an anomalous velocity along $b$ direction by $[\mathcal{F}_\lambda]_{n}^{ab}(t)$ for a band $n$~\cite{ishizuka2016emergent,jia2024equivalence,feng2025intrinsic}. If we write it in a more compact form, we have an equivalent expression of $[\mathcal{F}_\lambda]_{n}^{ab}(t)$ as
\begin{equation}
[\mathcal{F}_\lambda]_{n}^{ab}(t)=2\mathrm{Im}\langle\partial_b n|\partial_t n\rangle,
\end{equation}
where $|n\rangle \equiv |n(\lambda_a,t)\rangle$ is the eigenstate of the time-dependent Hamiltonian. This mixed Berry curvature is defined in momentum-time space and vanishes identically in the DC limit ($\Omega = 0$). Given the modified distribution function and anomalous velocity, one can directly obtain the first-order AC current. Specifically, we can write the first-order current as $J_b^{(1)}=(\sigma^{ab}_{\rm in}+\sigma^{ab}_{\mathrm{ex}})\lambda^a$ with $\sigma^{ab}_{\rm in}$ the intrinsic and $\sigma^{ab}_{\rm ex}$ the extrinsic conductivity. They can be derived as
\begin{equation}
\sigma^{ab}_{\mathrm{ex}}=-e \mathrm{Re}(\frac{\Omega \tau}{\Omega\tau+i}e^{i\Omega t}) \sum_n\int_{\bm{k}}v_n^b\mathcal{O}_{n}^a f'_n
\end{equation}
and 
\begin{equation}
\sigma^{ab}_{\rm in}=-e\sum_n\int_{\bm{k}}[\mathcal{F}_\lambda]_{n}^{ab}(t)f_n.
\end{equation}

We clarify that \(\sigma^{ab}_{\mathrm{in}}\) and \(\sigma^{ab}_{\mathrm{ex}}\) arise from distinct physical processes. \(\sigma^{ab}_{\mathrm{ex}}\) originates from scattering and contains a factor that depends on \(\Omega \tau\); this contribution is consistent with the extrinsic gyrotropic magnetic effect~\cite{zhong2016gyrotropic}. In contrast, \(\sigma^{ab}_{\mathrm{in}}\) is intrinsic and independent of scattering, corresponding to the intrinsic gyrotropic magnetic current~\cite{xiang2025intrinsic}. These two conductivities may also obey different symmetry constraints. It is also important to note that any energy correction to the band energy \(\varepsilon_n(\bm{k},t)\) is compensated in the instantaneous distribution function \(f[\tilde{\varepsilon}_n(\bm{k},t)] = f'_n[\tilde{\varepsilon}_n(\bm{k},t) - \varepsilon_n(\bm{k})]\). Consequently, given that both \(\sigma^{ab}_{\mathrm{in}}\) and \(\sigma^{ab}_{\mathrm{ex}}\) are purely dynamical in origin, one can verify that the resulting current vanishes identically in the DC limit (\(\Omega \rightarrow 0\)), as expected from physical intuition.

The first-order response to an AC drive oscillates at the drive frequency and vanishes upon time averaging, producing no net charge transport over a complete cycle. In contrast, the second-order nonlinear response can generate a time-independent DC current. This rectified current is essential for applications such as energy harvesting, wireless charging, and signal detection~\cite{onishi2024high}. Expanding the current density to second order in the drive amplitude, we obtain
\begin{equation}
J^{(2)}(t) = J_{\mathrm{dc}} + \mathrm{Re}\left[J_{2\Omega}\, e^{-2i\Omega t}\right]
\end{equation}
where the DC component $J_{\mathrm{dc}}$ is the rectified current central to this work, and $J_{2\Omega}$ describes second-harmonic generation. In what follows, we focus on the rectification contribution, which arises from the dynamical breaking of time-translation invariance and has no analog in the static limit. The currents for second-harmonic generation are present in Appendix B.

To obtain a solution corresponding to a steady-state DC current, we expand both the time-dependent velocity \( v_{n}(\bm{k}, t) \) and the distribution function \( f_n(\bm{k}, t) \) in harmonics of the drive frequency. The rectification of an alternating field can be understood as a two-wave mixing process, in the sense that the nonlinear interaction of the field with itself generates a time-independent component. Within this framework, we derive both the extrinsic and intrinsic contributions to the rectified current $J^c_{\mathrm{dc}}=(\sigma_{\mathrm{ex}}^{abc}+\sigma_{\mathrm{in}}^{abc})\lambda^a\lambda^b$, with the conductivities can be derived as follows
\begin{align}
&\sigma_{\mathrm{ex},1}^{abc}=e\frac{(\Omega \tau)^2}{1+(\Omega \tau)^2}\int_{n\bm{k}}\partial_c(\mathcal{O}_{n}^a\mathcal{O}_{n}^b)f'_n,\label{eq:curr1}\\
&\sigma_{\mathrm{ex},2}^{abc}=-\frac{4e\Omega^2 \tau}{1+(\Omega \tau)^2}\int_{n\bm{k}}\sum_{m\neq n}\frac{\mathrm{Im}(v_{mn}^c \mathcal{O}_{nm}^a)\mathcal{O}_{n}^b}{\varepsilon_{nm}^2}f'_n,\label{eq:curr2}\\
&\sigma_{\mathrm{in}}^{abc}=-4e\Omega^2 \int_{n\bm{k}}\sum_{m\neq n}\frac{\mathrm{Re}(v_{mn}^c\mathcal{O}_{nm}^a )\mathcal{O}_{n}^b}{\varepsilon_{nm}^3}f'_n.\label{eq:curr3}
\end{align}

\begin{figure}
\includegraphics[width=0.95\linewidth]{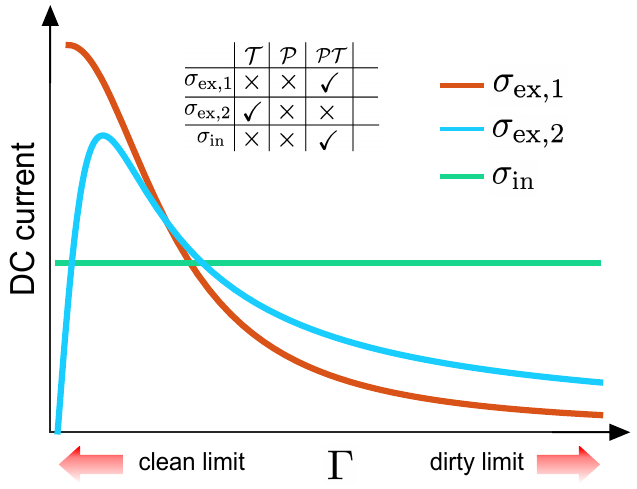}
\caption{Schematic depiction of the extrinsic ($\sigma_{\mathrm{ex},1}$ and $\sigma_{\mathrm{ex},2}$) and intrinsic ($\sigma_{\mathrm{in}}$) DC current and their symmetry properties. $\Gamma$ denotes the scattering rate, making $\sigma_{\mathrm{ex},1}$ term dominate and suppressing $\sigma_{\mathrm{ex},2}$ in the clean limit. However, the intrinsic one is disorder free and may dominate gradually in the dirty limit.}
\label{fig:fig2}
\end{figure}

These three equations are the main result of this work. The three conductivities \(\sigma_{\mathrm{ex,1}}, \sigma_{\mathrm{ex,2}}\) and \(\sigma_{\mathrm{in}}\) exhibit distinct scaling laws with the scattering rate \(\Gamma\) and are subject to different symmetry constraints. $\sigma_{\mathrm{ex,2}}$ and \(\sigma_{\mathrm{in}}\) are contributed from the mixed Berry curvature $[\mathcal{F}_\lambda]_{n}^{ab}(t)$ in Eq.~\eqref{eq:eq_mbc}. As shown in Fig.~\ref{fig:fig2}, \(\sigma_{\mathrm{ex,1}}\) denotes the reactive current saturates to a constant value in the high-frequency regime (clean limit, \(\Omega \gg \Gamma\)) and is suppressed as \(\sim 1/\Gamma^2\) in the low-frequency regime (\(\Omega \ll \Gamma\)). In contrast, the dissipative response \(\sigma_{\mathrm{ex,2}}\) vanishes for \(\Omega \gg \Gamma\) and decays as \(\sim 1/\Gamma\) when \(\Omega \ll \Gamma\). More importantly, the intrinsic current \(\sigma_{\mathrm{in}}\) is independent of \(\Gamma\). All three currents require broken inversion ($\mathcal{P}$) symmetry. However, \(\sigma_{\mathrm{ex,1}}\) and \(\sigma_{\mathrm{in}}\) are compatible with \(\mathcal{PT}\) (combined space-time) symmetry, whereas \(\sigma_{\mathrm{ex,2}}\) can survive when time-reversal ($\mathcal{T}$) symmetry is preserved.


Despite the driving frequency lying below the interband transition threshold, we find that both the extrinsic and intrinsic currents persist and exhibit a Fermi-surface nature. The intrinsic, disorder-free rectified current, $\sigma_{\mathrm{in}}$, acts as an ideal, reversible, and dissipationless energy conveyor, analogous to the nonlinear Hall effect~\cite{sodemann2015quantum,shi2023berry,matsyshyn2023fermi,onishi2022effects}. While the extrinsic current $\sigma_{\mathrm{ex,1}}$ saturates to a finite value in the clean limit, it nonetheless generates Joule heating due to its reliance on impurity scattering. The second extrinsic contribution, $\sigma_{\mathrm{ex,2}}$, arises from both the mixed Berry curvature and scattering processes and is nonmonotonic in the scattering rate $\Gamma$. Together, these results establish a unified framework for dynamical current generation.

\begin{figure*}
\includegraphics[width=1\linewidth]{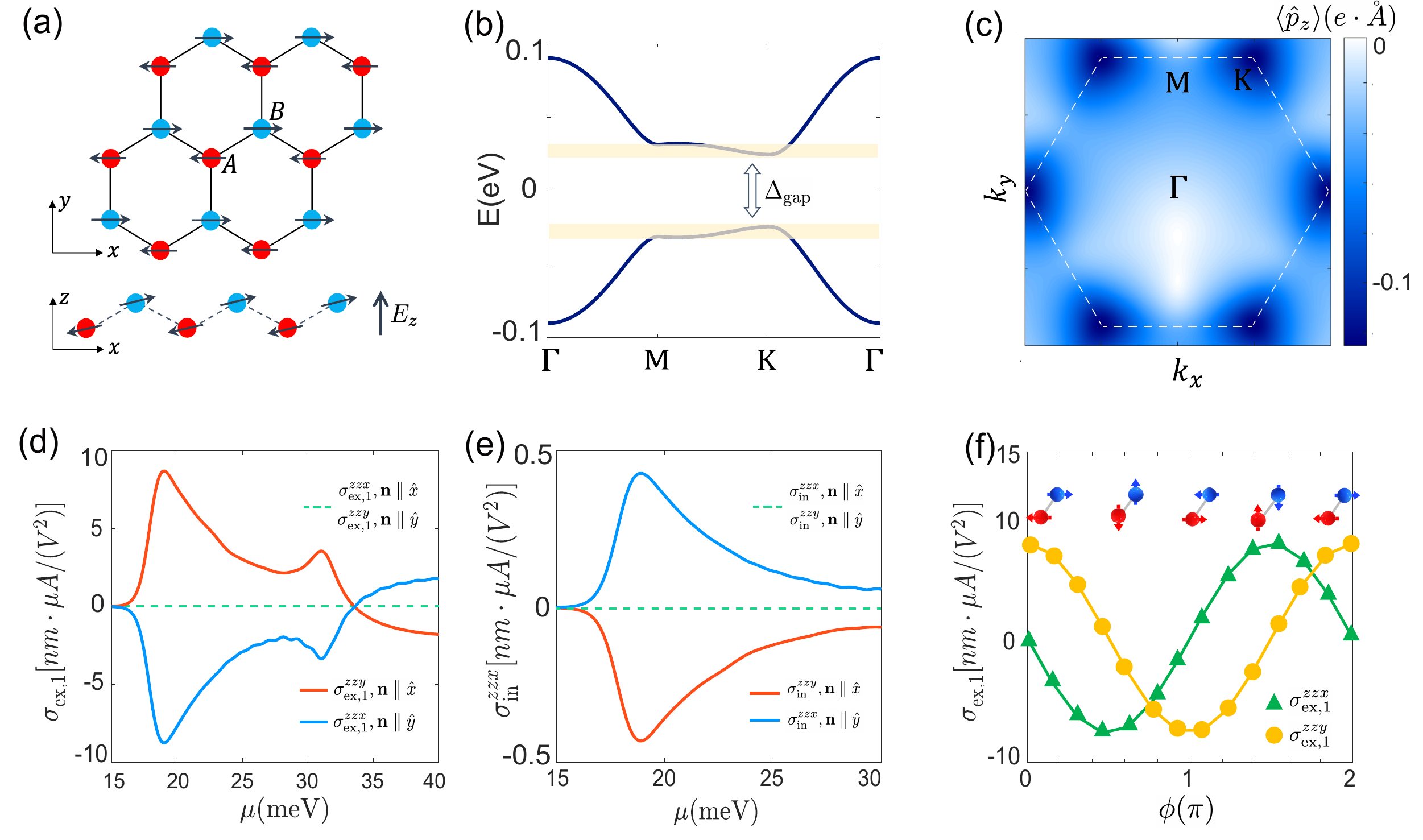}
\caption{(a) Schematic illustration of the buckled honeycomb lattice with N\'{e}el order. In the side view, the two sublattice atoms (A and B) form a vertical dipole moment that couples to an out-of-plane electric field $E_z$. (b) Band structure along the high-symmetry lines, featuring a band gap of $\sim 50$ meV. (c) Momentum-space distribution of the vertical dipole moment $\langle \hat{p}_z\rangle$ for the valence band. (d) and (e) Calculated $\sigma_{\mathrm{ex},1}$ and $\sigma_{\mathrm{in}}$ as functions of Fermi energy $\mu$. (f) Angular dependence of $\sigma_{\mathrm{ex},1}$ at $\mu=18$ meV on the N\'{e}el vector direction; the evolution of the spin texture on the A and B sublattices is also depicted. Parameters used: $t = 30$ meV, $\lambda_R = 0.1t$, $J_{ex} = 0.3t$, and temperature $T = 5$ K.}
\label{fig:fig3}
\end{figure*}


\emph{Nonlinear magnetoelectric gyrotropic current.}---Having established a general framework for the dynamical generation of nonlinear currents, we now seek a feasible physical realization. Concretely, as illustrated in Fig.~\ref{fig:fig3}(a), we consider a honeycomb lattice with a buckled magnetic ordering subjected to an out-of-plane oscillating electric field. The electric field $E_z$ couples to the out-of-plane electric dipole moment via
\begin{equation}
H' = -\hat{p}_z E_z(t),
\end{equation}
where $\hat{p}_z$ is the out-of-plane dipole moment operator. This coupling arises exclusively from the internal sublattice asymmetry inherent to the buckled structure. Such an electric field effectively mimics a magnetic field, enabling us to explore gyrotropic-like currents without applying a dynamical magnetic field~\cite{zhang2024predictable}. It is also important to note that in equilibrium, no net current is generated. Therefore, we can take $\hat{\mathcal{O}} \rightarrow -\hat{p}_z$ and $\lambda(t) \rightarrow E_z(t)$ to evaluate the rectified current response.

Next we take the tight binding model with noncollinear N\'{e}el ordering~\cite{sivadas2016gate,ren2025nonreciprocal}, which reads
\begin{align}
H =  &-t \sum_{\langle ij \rangle} c_i^{\dagger}c_j 
 - i\lambda_{R} \sum_{\langle\langle ij \rangle\rangle } \mu_{ij} c_i^\dagger (\bm{s}\times \hat{\bm{d}}_{ij})_z c_j  \\\nonumber
 &+J_{ex}\sum_{i}(-1)^i c_i^{\dagger}(\bm{n}\cdot\bm{s})c_i,
\end{align}
where the first term \(t\) represents the nearest-neighbor hopping amplitude. The second term describes the intrinsic Rashba spin-orbit coupling, with \(\mu_{ij} = \pm 1\) for \(\{i,j\}\) belonging to the A or B sublattice. The last term represents a staggered exchange field arising from the N\'{e}el order. \(\bm{s} = (s_x, s_y, s_z)\) are the Pauli matrices in spin space. Here, \(J_{\mathrm{ex}}\) is the exchange coupling strength, and \(\bm{n}\) is the N\'{e}el order vector. As a model study, we take $\lambda_R=0.1t$ and $J_{ex}=0.3t$. The band structure with $t=30$ meV is shown in Fig.~\ref{fig:fig3}(b) and a band gap of about 50 meV can be seen.

\begin{table}
\caption{Crystalline symmetry restrictions of nonlinear magnetoelectric current for the $\sigma^{zzx}$ and $\sigma^{zzy}$. $\checkmark$ indicates allowed and $\times$ indicates forbidden.}
\label{tab:summary}
\begin{ruledtabular}
\begin{tabular}{l|llll}
 & $\mathcal{P}$ & $C_{nz}$ & $\mathcal{M}_x/C_{2x}$ & $\mathcal{M}_y/C_{2y}$  \\
 \hline
$\sigma^{zzx}$ & $\times$ & $\times$     & $\times$ &   \checkmark  \\
$\sigma^{zzy}$ & $\times$ & $\times$  &  \checkmark  &   $\times$ 
\end{tabular}
\end{ruledtabular}
\end{table}

Guided by the $\mathcal{PT}$ symmetry of this model Hamiltonian, we focus on $\sigma_{\mathrm{ex},1}$ and $\sigma_{\mathrm{in}}$ introduced in the previous section. Table.~\ref{tab:summary} presents a symmetry classification of the two conductivity components $\sigma^{zzx}$ and $\sigma^{zzy}$ (for brevity we omit the subscript of $\sigma$). Figure.~\ref{fig:fig3}(c) shows the momentum-space distribution of the vertical dipole moment $\langle \hat{p}_z\rangle$ for the valence band with $\bm{n}\parallel\hat{x}$; the distribution for the other degenerate band follows from $\mathcal{PT}$ operation. Figure.~\ref{fig:fig3}(d) displays $\sigma_{\mathrm{ex},1}$ in the clean limit as a function of Fermi energy $\mu$ for N\'{e}el vector $\bm{n}\parallel\hat{x}$ and $\hat{y}$. When $\bm{n}\parallel\hat{x}$ ($\bm{n}\parallel\hat{y}$), mirror symmetry $\mathcal{M}_x$ ($\mathcal{M}_y$) forbids the current along $x$ ($y$). The intrinsic response $\sigma_{\mathrm{in}}$ at $\Omega = 1$ THZ exhibits similar behavior, as shown in Fig.~\ref{fig:fig3}(e), although its magnitude is smaller due to the $1/\Delta_{\mathrm{gap}}^3$ scaling. In real materials, a finite scattering rate $\Gamma$ suppresses $\sigma_{\mathrm{ex},1}$, allowing the two contributions to become comparable in magnitude.

Perhaps more interesting, the rectified current depends sensitively on the N\'{e}el ordering, making it a powerful tool for detecting the N\'{e}el vector in buckled magnetic materials. The N\'{e}el vector can be parameterized as $\bm{n} = (\sin\theta\cos\phi,\sin\theta\sin\phi,\cos\theta)$ (we take $\theta = \pi/2$ for in-plane ordering), yielding $\sigma_{\mathrm{ex},1}^{zzx} \propto \sin\phi$ and $\sigma_{\mathrm{ex},1}^{zzy} \propto \cos\phi$, as shown in Fig.~\ref{fig:fig3}(f). The vanishing of these components at specific angles can be understood from mirror symmetry: when $\bm{n}$ aligns perpendicular to a mirror plane, the corresponding mirror symmetry strictly forbids the longitudinal current response. This results in a $2\pi$ periodicity in the angular dependence.

While high-frequency magnetic fields are challenging to realize because of the inductive impedance of coils and the scarcity of efficient sources in the terahertz regime, oscillating electric fields can be readily generated using standard optical or terahertz techniques~\cite{chen2024crossed,andrei2020graphene,wilson2021excitons,regan2022emerging,drexler2013magnetic}. Taking a characteristic value of conductivity $\sigma=1$ $\rm{nm \cdot\mu A/V^2}$ and electric field $10$ mV/nm, the current density is about $0.1$ $\rm{\mu A/\mu m}$, which is a sizable value to be measured in experiments.

\emph{Conclusion and discussion.}---We have developed a general framework for the dynamical generation of rectified electric currents driven by an external perturbation that has no static counterpart. Importantly, we have identified two distinct origins of this dynamical rectification, encapsulated in three current equations [Eq.~\eqref{eq:curr1} to \eqref{eq:curr3}]. The extrinsic contribution arises from impurity scattering at finite frequency; this process transfers energy to the crystal lattice in the form of heat. The intrinsic contribution, in contrast, generates a rectified current that flows frictionlessly into the output circuit. Such intrinsic responses, rooted in band quantum geometry, have garnered significant interest in recent years~\cite{xiao2022intrinsic,huang2023intrinsic,wang2024orbital,wang2025intrinsic,zheng2024interlayer,zhang2025intrinsic,xiang2025spin,das2023intrinsic,wei2023quantum,qiang2026quantum}.

Detecting the orientation of the N\'{e}el vector is a central task in antiferromagnetic spintronics. The predicted nonlinear magnetoelectric gyrptropic current is generic to a broad class of buckled magnetic materials~\cite{liu2026symmetry}. Additionally, while we have employed a simple model to treat impurity scattering, more exotic mechanisms, such as skew scattering, present interesting directions for future study~\cite{gong2025identifying,ma2023anomalous,du2019disorder,du2021quantum,atencia2023disorder,nandy2019symmetry,xiao2019theory}. Our work also motivates more realistic calculations in concrete materials with complex band structures using the formulas developed here.

\emph{Acknowledgments.}---We thank Likun Shi for helpful discussions. J.X.H. is supported by the postdoctoral fellowship of Hong Kong University of Science and Technology. C.W. is supported by the National Natural Science Foundation of China under the Grants No. 12234016 and No. 12174317. This work has been supported by the New Cornerstone Science Foundation.


%

\clearpage
\bigskip
\onecolumngrid
\begin{center}\textbf{End Matter}\end{center}
\twocolumngrid
\vspace{0.5\baselineskip}

\appendix
\renewcommand{\theequation}{A\arabic{equation}} 
\setcounter{equation}{0} 
\renewcommand{\thefigure}{A\arabic{figure}}
\setcounter{figure}{0}

{\emph{Appendix A: Time-dependent Schrieffer-Wolff transformation.}}
For a system governed by a time-dependent Hamiltonian, one can find a unitary transformation that rotates the system into a frame in which the transformed Hamiltonian is block-diagonal to a desired order. The generator $S$ is typically chosen such that the first-order correction to $H_0$ cancels the unperturbed Hamiltonian at zeroth order, i.e. $H\rightarrow e^{S(t)}H e^{-S(t)}-i e^{S(t)}\partial_t e^{-S(t)}$. This gives the equation
\begin{equation}
[S,H_0]+i\partial_t S+\hat{\bm{O}}\cdot \bm{\lambda}(t)=0.
\end{equation}
The generator $S$ can be solved as
\begin{equation}
S_{mn}(t)=\lambda_a \mathcal{O}_{mn}^a(\frac{e^{-i\Omega t}}{\varepsilon_{mn}-\Omega}+\frac{e^{i\Omega t}}{\varepsilon_{nm}+\Omega}).
\end{equation}
Therefore, the first order perturbation of velocity is
\begin{equation}
\delta v_{n}^b(t)=[S,v^b]_{nn}=\lambda^a(t)\partial_b\mathcal{O}_{n}^a+\lambda^a [\mathcal{F}_\lambda]_{n}^{ab}(t)
\end{equation}
Here, the first term is the velocity from band energy correction, while the second term denotes the anomalous velocity. $[\mathcal{F}_\lambda]_{n}^{ab}(t)$ is the dynamical mixed Berry curvature given in the main text. On the other hand, we note that the time-dependent wave function follows
\begin{equation}
|n(\lambda,t)\rangle=|n\rangle+\sum_{\substack{m\neq n\\ \omega=\pm\Omega}}\lambda^a \mathcal{O}_{mn}^a\frac{e^{i\omega t}}{\varepsilon_{nm}+\omega}|m\rangle.
\end{equation}
In this way, the $[\mathcal{F}_\lambda]_{n}^{ab}(t)$ can also be written as $[\mathcal{F}_\lambda]_{n}^{ab}(t)=2\mathrm{Im}\langle\partial_b n(\lambda,t)|\partial_t n(\lambda,t)\rangle$.

\renewcommand{\theequation}{B\arabic{equation}} 
\setcounter{equation}{0} 
\renewcommand{\thefigure}{B\arabic{figure}}
\setcounter{figure}{0}

{\emph{Appendix B: Second harmonic generation by dynamical drive.}}
The rectified currents have been analyzed in the main text. For completeness, here we present the corresponding second harmonic generation (SHG) currents. We identify four distinct SHG contributions: two extrinsic contributions that depend explicitly on the scattering rate, and two intrinsic contributions that are independent of disorder. These can be expressed as follows
\begin{align}
&\sigma_{\mathrm{ex},1}^{abc}=e\mathrm{Re}[(\frac{2\Omega \tau}{-2\Omega\tau+i}-\frac{\Omega \tau}{-\Omega\tau+i})e^{2i\Omega t}]\int_{n\bm{k}}\partial_c(\mathcal{O}_{n}^a\mathcal{O}_{n}^b)f'_n\\
&\sigma_{\mathrm{ex},2}^{abc}=\frac{2e\Omega^2 \tau \cos(2\Omega t)}{1+(\Omega \tau)^2}\int_{n\bm{k}}\sum_{m\neq n}\frac{\mathrm{Im}(v_{mn}^c \mathcal{O}_{nm}^a)\mathcal{O}_{n}^b}{\varepsilon_{nm}^2}f'_n\\
&\sigma_{\mathrm{in},1}^{abc}=-2e\Omega  \sin(2\Omega t)\int_{n\bm{k}}\sum_{m\neq n}\frac{\mathrm{Im}(v_{mn}^c \mathcal{O}_{nm}^a)\mathcal{O}_{n}^b}{\varepsilon_{nm}^2}f'_n\\
&\sigma_{\mathrm{in},2}^{abc}=-2e\Omega^2  \cos(2\Omega t)\int_{n\bm{k}}\sum_{m\neq n}\frac{\mathrm{Re}(v_{mn}^c \mathcal{O}_{nm}^a)\mathcal{O}_{n}^b}{\varepsilon_{nm}^3}f'_n.
\end{align}
After the symmetry analysis, $\sigma_{\mathrm{ex},1}^{abc}$ and $\sigma_{\mathrm{in},2}^{abc}$ are time-reversal odd, while $\sigma_{\mathrm{ex},2}^{abc}$ and $\sigma_{\mathrm{in},1}^{abc}$ are time-reversal even. Therefore, for $\mathcal{PT}$ symmetric system studied in the main text, we expect the SHG currents $\sigma_{\mathrm{ex},1}^{abc}$ and $\sigma_{\mathrm{in},2}^{abc}$ are nonzero.
\renewcommand{\theequation}{D\arabic{equation}}
\setcounter{equation}{0}

\end{document}